\documentclass[sigconf, pbalance=true]{acmart}
\AtBeginDocument{%
  }

\setcopyright{acmlicensed}
\copyrightyear{2025}
\acmYear{2025}

\acmConference[PROBabLE Futures Interim Report]{Probabilistic AI Systems in Law Enforcement Futures}{December, 2025}{RAi Keystone Project}
\usepackage{subfig}
\usepackage{makecell}
\usepackage{multirow}
\usepackage{colortbl}
\usepackage{color, soul}
\usepackage{tabstackengine}

\usepackage{xcolor}
\usepackage[most]{tcolorbox}

\usepackage{todonotes}
\usepackage{url}

\begin{document}

\title{Mapping the Probabilistic AI Ecosystem in Criminal Justice in England and Wales - Interim Report}%Probabilistic AI in Criminal Justice: Mapping the Ecosystem and Scoping for the Future

\author{Evdoxia Taka}
\affiliation{%
  \institution{University of Glasgow}
  \city{Glasgow}
  \country{UK}}
\email{evdoxia.taka@glasgow.ac.uk}

\author{Temitope Lawal}
\affiliation{%
  \institution{University of Northumbria}
  \city{Newcastle}
  \country{UK}}
\email{temi.lawal@northumbria.ac.uk}

% \author{Katherine Jones}
% \affiliation{%
%   \institution{University of Northumbria}
%   \city{Newcastle}
%   \country{UK}}
% \email{katherine.jones@northumbria.ac.uk}

\author{Muffy Calder}
\affiliation{%
  \institution{University of Glasgow}
  \city{Glasgow}
  \country{UK}}
\email{muffy.calder@glasgow.ac.uk}

\author{Michele Sevegnani}
\affiliation{%
  \institution{University of Glasgow}
  \city{Glasgow}
  \country{UK}}
\email{michele.sevegnani@glasgow.ac.uk}

\author{Kyriakos Kotsoglou}
\affiliation{%
  \institution{University of Northumbria}
  \city{Newcastle}
  \country{UK}}
\email{kyriakos.kotsoglou@northumbria.ac.uk}

\author{Elizabeth McClory-Tiarks}
\affiliation{%
  \institution{Newcastle University}
  \city{Newcastle Upon Tyne}
  \country{UK}}
\email{lizzie.tiarks@newcastle.ac.uk}

\author{Marion Oswald}
\affiliation{%
  \institution{University of Northumbria}
  \city{Newcastle}
  \country{UK}}
\email{marion.oswald@northumbria.ac.uk}

\renewcommand{\shortauthors}{Taka et al.}

 
\begin{abstract}
Commercial or in-house developments of probabilistic AI systems are introduced in policing and the wider criminal justice (CJ) system worldwide, often on a force-by-force basis. We developed a systematic way to characterise probabilistic AI tools across the CJ stages in a form of mapping with the aim to provide a coherent presentation of the probabilistic AI ecosystem in CJ. We use the CJ system in England and Wales as a paradigm. This map will help us better understand the extent of AI's usage in this domain (how, when, and by whom), its purpose and potential benefits, its impact on people's lives, compare tools, and identify caveats (bias, obscured or misinterpreted probabilistic outputs, cumulative effects by AI systems feeding each other, and breaches in the protection of sensitive data), as well as opportunities for future implementations. In this paper we present our methodology for systematically mapping the probabilistic AI tools in CJ stages and characterising them based on the modes of data consumption or production. We also explain how we collect the data and present our initial findings. This research is ongoing and we are engaging with UK Police organisations, and government and legal bodies. Our findings so far suggest a strong reliance on private sector providers, and that there is a growing interest in generative technologies and specifically Large Language Models (LLMs).
\end{abstract}



% \centering
%   \includegraphics[width=0.7\textwidth]{mapping}
%   \caption{AI tools currently used in the \emph{Intelligence} stage of the Criminal Justice (CJ) system in England and Wales. The purpose and taxonomy of each tool are stated below the name of each tool. The colour of each tool's box indicates the implementation stage of the tool: green for deployed, yellow for trialed, and pink for experimental. Shaded boxes indicate tools not used in the corresponding stages of CJ.}
%   \Description{The figure presents 20 AI tools used in the Intelligence stage of the Criminal Justice process in a 5x4 grid.}
%   \label{fig:mapping}

%\received{30 April 2025}
% \received[revised]{21 July 2025}
%\received[accepted]{2 July 2025}

\maketitle
% \today

\section{Introduction}

Probabilistic AI systems, such as facial recognition, risk predictive tools, and large language models (LLMs) are introduced in policing and the wider criminal justice (CJ) system   in many countries worldwide, including the UK. There are a variety of implementations, commercial and in-house, and applications and variations across countries. These tools are also very often introduced, implemented, trialled, and deployed on a force-by-force basis~\cite{Oswald2018}.

Mapping the probabilistic AI ecosystem in CJ systematically could help us understand the extent of AI usage (how, when, and by whom) and its impact on people's lives, compare tools, and identify caveats (bias~\cite{Oswald2025,Oswald2018}, obscured or misinterpreted probabilistic outputs~\cite{Kotsoglou2020,Oswald2018decmak}, cumulative effects by AI systems feeding each other~\cite{Janjeva2023}, and breaches in the protection of sensitive data), as well as lessons from current deployments and opportunities for future implementations. Such  a mapping would help us to identify implications for other organizations in the CJ system and enable organizations to consider how AI outputs may feed into evidential contexts. It could also help  to identify what is required to build a tested and trusted framework to steer the development and deployment of AI tools in CJ so  they are responsible, effective, and trusted by the communities  and they respect their  rights.

We developed a systematic way to characterise probabilistic AI tools across the CJ stages with the aim of providing a coherent presentation of the AI-tools-in-CJ landscape. We use the paradigm of the England and Wales CJ system and have examples of real AI tools currently in use. Both legal and technical perspectives are considered to create this map. Although this mapping is reflective of AI tools in use in the England and Wales' CJ system, it is nonetheless easily adaptable to ecosystems in other countries as the methodological underpinnings are applicable irrespective of jurisdictional peculiarities.

This mapping is a holistic approach to explore the application of AI in CJ and better understand where and how it is used in CJ. This will be beneficial for potential stakeholders, e.g., police officers, lawyers, judges, commercial system developers, and public sector bodies, but also for the general public as it will provide a deeper understanding of the extent to which the AI tools used in CJ do and will affect our lives. 

%Our challenge is to acquire the necessary data to develop this mapping, but this research is ongoing. 

%\mscom{We can integrate below the paragraph on AI vs non-AI tools}
We relied on publicly available information and conducted semi-structured interviews with UK Police organizations, and government and legal bodies to acquire the necessary data to develop this mapping. Our purpose in this paper is to present the methodology we followed to develop the mapping (how we split the CJ system in England and Wales into stages, how we taxonomise  AI tools based on the modes of inputs-outputs and inference, what types of data we use and how we collected it) and present our initial findings. Overall our contributions are:
%\mscom{Some refs to semi-structured interview methods \url{https://www.maxqda.com/research-guides/semi-structured-interviews}}

\begin{itemize}
    \item a methodology to map the probabilistic AI ecosystem in Criminal Justice,
    \item a data repository of identified AI tools used, trialled or in experimental stage across the CJ system in England and Wales,
    \item useful insights about the breadth of AI use across  the CJ system in England and Wales based on this data. 
\end{itemize}

\section{Methodology}

\subsection{Staging of the CJ cystem in England \& Wales}

The CJ system is not uniform across the UK. There are three distinct systems found in this territory, one in England and Wales, another in Scotland, and a separate one in Northern Ireland. We focus on England and Wales, where the majority of police forces is found. Police forces in the UK are independent bodies and they are broken down by territory (from the 45 territorial forces, 39 are in England, 4 in Wales, 1 in Scotland, and 1 in Northern Ireland).  %There are also three specialist national forces: the British Transport Police, the Ministry of Defence Police, and the Civil Nuclear Constabulary.  

We divide the CJ system in England and Wales into stages based on its main functions and purposes, as follows:

\vspace{1mm}
\noindent\textbf{Stage 1: Community Policing and Offender Management.} This stage involves the police co-operating with community groups and multi-agency approaches targeted at tackling persistent offenders and/or offenders assessed as being at high risk of reoffending.
%focuses on preventing crime before it happens through engagement with the community, problem-solving initiatives, and close monitoring of offenders. 
%Police work alongside local organisations, schools, and community leaders to build trust and improve safety.  

\vspace{1mm}
\noindent\textbf{Stage 2: Intelligence.} This is the use of information given to or collected by the police  
%gathers and analyses information
 to initiate or support ongoing investigations.
%to detect potential criminal activity. 
%Intelligence tools, such as surveillance systems, 
% !! Muffy - we cant say surveillance system! 
%inform decision-making and crime prevention strategies.  

\vspace{1mm}
\noindent\textbf{Stage 3: Investigation.}
This is where police collect and consider evidence to investigate an alleged offence, e.g. collecting witness statements, interviewing  suspects, considering any digital (e.g. data from computers or phones) or forensic evidence (e.g. DNA).
%Once a crime has been reported or detected, law enforcement investigates to gather evidence, interview witnesses, and identify suspects. Advanced forensic techniques, cyber-investigations, and AI-based analysis are often used.  

\vspace{1mm}
\noindent\textbf{Stage 4: Charging Decision or Alternative Disposal.} The police can make the decision to charge a suspect, or offer an alternative disposal, such as a caution or penalty notice, for less serious offences. The Crown Prosecution Service (CPS) makes the charging decision for more serious offences, which involves an assessment of whether there is enough admissible and reliable evidence for a realistic prospect of conviction; and whether it is in the public interest to prosecute.

\vspace{1mm}
\noindent\textbf{Stage 5: Trial or Guilty Plea.} If the defendant pleads not guilty, a trial will take place in the magistrates' court or Crown Court, depending on the seriousness of the offence. If the defendant pleads guilty, the case will progress to the sentencing stage.

%or handled through an out-of-court settlement, such as fines or community service for minor offences.  

\vspace{1mm}
\noindent\textbf{Stage 6: Sentencing.} If an offender pleads, or is found guilty, they are sentenced. There are a wide range of disposals (and differences in availability of disposals between adult and young offenders), e.g. a fine, community sentence or a custodial sentence.  

%\etcom{We assume here that people know what probation is. Is it a service? Could we provide a short explanation? Lizzie or Kyri?}
%\mccom{I dont think we need to define probation, beyond what we have written for stage 8}
\vspace{1mm}
\noindent\textbf{Stage 7: Prison and Parole.} For serious crimes, offenders can be sentenced to imprisonment. When offenders are released on parole part-way through their custodial sentence, this is supervised by probation.  
%allows them to serve the remainder of their sentence under supervision.  

\vspace{1mm}
\noindent\textbf{Stage 8: Probation.} Individuals who receive a community order involving e.g. unpaid work or electronic monitoring, are managed and supervised by probation.
%After serving their sentence, some offenders are placed on probation to ensure they reintegrate into society and do not reoffend. This may involve mandatory check-ins, employment requirements, or restrictions on certain activities. 

\vspace{1mm}
AI tools may be deployed in more than one CJ stage; 
mapping  tools across the stages enables us  to analyse and understand the breadth of AI use.   For example, it could help us identify areas of heavy use of AI and inform future investigations concerned with potential risks and existence or absence of sufficient policies and regulations. It could also indicate potential ways to consolidate, thus achieving additional consistency and productivity.

\subsection{Taxonomy of AI tools}
%\mccom{Made changes in this section. I have tried to distinguish between the   taxonomy, which  provides a way to classify tools, and the   result, which  is a classification.}
 Rather than categorising AI tools by algorithm type or ML models, we developed a \emph{taxonomy} that  distinguishes non-generative and generative AI, with a further distinction within the non-generative class.
 This serves as a more precise and common vocabulary when mapping the landscape, enabling us to compare tools' capabilities, with respect to their application, not internal workings, and  assess relative popularity.  It is  based on  the mode in which information is inferred from inputs (\emph{inference modes}) and the mode of inputs and outputs (\emph{input-output modes}). 
 %\mscom{Other taxonomies we can check: \url{https://www.tuev-lab.ai/fileadmin/user_upload/TUEV_AI_Lab_Whitepaper_TUEV_AI_System_Taxonomy_EN.pdf}, \url{https://deepwiki.com/joylarkin/2025-AI-Topics/7.1-generative-and-multimodal-ai}, \url{https://github.com/danielrosehill/Multimodal-AI-Taxonomy}, \url{https://nvlpubs.nist.gov/nistpubs/ai/NIST.AI.200-1.pdf}, \url{https://www.eit.europa.eu/sites/default/files/creation_of_a_taxonomy_for_the_european_ai_ecosystem_final.pdf}.}

\vspace{1mm}
\noindent We identify three \emph{inference modes}:

\vspace{1mm}
\noindent\textbf{Analysis} of content, e.g. classify, match, cluster, transcribe. Examples are when we recognise a face or voice, identify a weapon in an image, or transcribe from audio to text.

\vspace{1mm}
\noindent\textbf{Synthesis} and enrichment, e.g. combine datasets to predict, profile. Examples are when we profile the probability to re-offend, or predict spatio-temporal crime hot spots.

\vspace{1mm}
\noindent\textbf{Generation} of content, e.g. generate from a prompt or a prompt and text input. Examples are the summarisation of a statement, generation of questions for a witness interview, RAG (retrieval augmented generation).

\vspace{1mm}
\noindent We   identify six \emph{input-output modes}:

\vspace{1mm}
\noindent\textbf{Primary data}
\begin{enumerate}
    \item text
    \item image
    \item video
    \item audio
\end{enumerate}

\vspace{1mm}
\noindent\textbf{Structured or enumerated data}
\begin{enumerate}
    \item[(5)] enum. This type denotes enumerated types, e.g.  Yes/No, sentiments, semantic features, points to prove (the elements of an offence that must be proved beyond reasonable doubt to secure a conviction). For some tools we may have a generally understood name for enum, e.g. Bool, Emotion, Name – of people, of repository, etc., Index – indices for a document repository.
\end{enumerate}

\vspace{1mm}
\noindent\textbf{Instructions for a generative AI model}
\begin{enumerate}
\item[(6)] prompt. The set of instructions that impacts the result.
\end{enumerate}

% \vspace{1mm}
An input or output can be individual item (from list above), a product or a list of items. A product of inputs or outputs is several modes separated by a comma (“,”); e.g. text, image. A product of modes indicates the combination of different types of data that can be of the same mode. For example, we can have a product of the same mode, e.g. enum, enum (an alternative notation for this would be $enum^n$, where $n=2$) for combining yes/no data and sentiments. A list of inputs or outputs is a list of a single mode, using the notation “[..]”; e.g. [image]. A list indicates a collection of multiple instances of the same type of data (e.g. images of faces). For example, live facial recognition compares a single face image against a watchlist [image] – a list of faces.

\vspace{1mm}
\noindent Overall, a class in the taxonomy has the form: 

\vspace{1mm}
\texttt{Inf(Modes$_{in}$} $\rightarrow$ \texttt{Modes$_{out}$)}, where

\texttt{Inf} $\in\{Analysis,Synthesis,Generation\}$

$\texttt{Modes} = \texttt{Mode} |  \texttt{Mode, Modes}$

$\texttt{Mode} = \texttt{mode} | [\texttt{Mode}]$

\texttt{mode} $\in\{text,image,video,audio,enum,prompt\}$

\vspace{1mm}

%\mccom{Moved table 1 description to before table 2 description and  removed the use of "stage"  - confusing with CJS stage}

Sequencing AI tools or capabilities, i.e. cases where the outputs become inputs, can     be  described based on the taxonomy. Table~\ref{tab:seq_ex}  contains a few examples.

To demonstrate the expressive power of our taxonomy, we provide in Table~\ref{tab:tax_ex} some   example  classifications along with a   brief description, based on commonly encountered applications of AI in policing in the UK. 
\begin{table}[!h]
\caption{Examples of  sequencing of AI tools. Outputs from the first tool are inputs to the  second tool.}

\resizebox{\linewidth}{!}{%
\begin{tabular}{@{}rll@{}}
\toprule
\bf Sequence &
  \bf Classification (First tool) &
 \bf Classification (Second tool) \\ 

\midrule
 \makecell[r]{\textbf{Transcribe,}\\ \textbf{then translate}}&  Analysis (audio $\rightarrow$ text)& Analysis (text $\rightarrow$ text)  \\

 \hline
 
\makecell[r]{\textbf{Transcribe,}\\ \textbf{then summarize}}& Analysis (audio $\rightarrow$ text)&  Generation (prompt, text $\rightarrow$ text) \\

\hline

\makecell[r]{\textbf{Extract hate speech}\\ \textbf{topics, then use} \\ \textbf{in synthesis of}\\ \textbf{offender profiles}}& Analysis (text $\rightarrow$ enum)&  Synthesis (enum, ... $\rightarrow$ enum)\\

\bottomrule
\end{tabular}
}
\label{tab:seq_ex}
\end{table}

\begin{table*}[!h]
\centering
\caption{Taxonomy Examples.}
\label{tab:tax_ex}
% \resizebox{\linewidth}{!}{%
 \begin{tabular}{lll}

    \toprule
    
     Inference Mode & Classification &   Description\\
     
     \midrule
           
   \multirow{23}{*}{\textbf{Analysis}}
   & \cellcolor{gray!25}Analysis (audio $\rightarrow$ text)& \cellcolor{gray!25}Audio to text transcription\\ 
   
   &Analysis (audio, audio $\rightarrow$ enum)& \makecell[l]{Speaker recognition (voice recognition)\\Here enum is Bool (i.e., yes or no) denoting\\ whether the speaker in audio2 occurs in audio1}\\ 
   
   &\cellcolor{gray!25}Analysis (audio, [audio] $\rightarrow$ enum)&\multicolumn{1}{>{\columncolor{gray!25}}l}{\tabularCenterstack{l}{Speaker recognition (voice recognition)\\Here enum is Bool (i.e. yes or no) denoting\\ whether any of the speakers in [audio2] occur in audio1}}\\ 
   
   &Analysis (audio $\rightarrow$ [audio])&\makecell[l]{Speaker segmentation. An audio stream is\\ structured into a list of audio segments, segmented\\ at speaker change points}\\ 
   
   &\cellcolor{gray!25}Analysis (audio, [audio] $\rightarrow$ [audio, enum])& \multicolumn{1}{>{\columncolor{gray!25}}l}{\tabularCenterstack{l}{Speaker diarisation (segmentation and recognition)\\ given a list of speaker audio samples. Here, enum\\ refers to the ordering in speaker samples,\\ i.e. speaker labelling is implied}}\\
   
   &Analysis (audio, [audio, enum] $\rightarrow$ [audio, enum]) &\makecell[l]{Speaker diarisation (segmentation and recognition),\\ given a list of speaker audio samples and (explicit) labels.\\ Here, enum refers to the speaker labels  in both the samples \\ and the segments}\\ 
   
   &\cellcolor{gray!25}Analysis (image $\rightarrow$ enum)&\multicolumn{1}{>{\columncolor{gray!25}}l}{\tabularCenterstack{l}{Identification of weapon in a scene.\\ Here, enum is Bool (i.e. yes or no)}}\\ 
   
   &Analysis (text $\rightarrow$ text) &Translation from Welsh to English\\ 
   
   &\cellcolor{gray!25}Analysis (image, image $\rightarrow$ image)&\multicolumn{1}{>{\columncolor{gray!25}}l}{\tabularCenterstack{l}{Bystander removal by removing occurrences of\\ 2nd image from 1st image}}\\
   
    \hline
    
    \multirow{3}{*}{\textbf{Synthesis}} 
    & Synthesis (text $\rightarrow$ enum) & \makecell[l]{Synthesis of points to prove from a witness statement\\ Here, enum is e.g. 5 different points}\\ 
    & \cellcolor{gray!25}Synthesis (enum, enum $\rightarrow$ enum) & 
    \multicolumn{1}{>{\columncolor{gray!25}}l}{\tabularCenterstack{l}{Crime hot-spot prediction. Here, 1st enum\\ is weather forecast, 2nd is temporal spatial, 3rd is Bool}}\\ 
    
    \hline
    
    \multirow{6}{*}{\textbf{Generation}} 
    & Generation (prompt, text $\rightarrow$ text)&Summarization of text\\ 
    &\cellcolor{gray!25}Generation (prompt, text $\rightarrow$ text)& \cellcolor{gray!25}Generation of interview questions in style of 1st text\\ 
    &Generation (prompt, text $\rightarrow$ [text])&Generation of a list of interview questions in style of 1st text\\ 
    &\cellcolor{gray!25}Generation (prompt, text $\rightarrow$ image)&\cellcolor{gray!25}Generation of a suspect image from the witness statement\\ 
    &Generation (prompt, name $\rightarrow$ text, [index])&\makecell[l]{Generation of a response, with list of citations,\\ to the prompt referring to the (named) repository\\ (RAG example)}\\
    
    \bottomrule%
\end{tabular}%
% }
\end{table*}

%\etcom{I have a feeling that the taxonomy seems unlinked to what we try to do with the mapping. We probably need to provide more argumentation about why we need it and why we include it in the mapping. We mention that it could serve as a common vocabulary to map the landscape, compare the tools' capabilities, and assess the involved risks. We could provide some examples to illustrate this. But I have no idea what sort of. Something that could demonstrate the value of the taxonomy based on our map. }
%\mccom{I think I have addressed this now }

\subsection{Description of the mapping data}
%\mccom{I dont understand the purpose of this paragraph - and why it is here.  Moreover, is this exactly true for generative tools?  I would remove, or at least move to earlier}
%\etcom{Yes, this applies to Generation - LLMs are trained on data. Even ngrams rely on data and its statistics - not sure though if we could say that these learn from data, but in any case I don't think that we encountered any tools that rely on these models. I move this at the beginning. The purpose of this paragraph is to explain our requirements for a tool to be considered in our mapping - we exclude tools that do not use ML.}

%\mscom{Other systematic surveys: \url{https://link.springer.com/article/10.1007/s42001-025-00373-z}, \url{https://www.mdpi.com/2073-431X/12/12/255}, \url{https://www.justice.gov/olp/media/1381796/dl?inline}, \url{https://www.policingproject.org/ai-explained-articles/2024/9/6/how-policing-agencies-use-ai}, \url{https://www.interpol.int/en/How-we-work/Innovation/Artificial-Intelligence-Toolkit}}
Our purpose in this research is to create a repository of AI tools that are used, have been trialled or are in an experimental stage within the CJ system in England and Wales. Artificial Intelligence (AI) is a very broad term referring to any algorithmic approach used to mimic human behavior. We only consider tools that use \emph{learning from data} through \emph{Machine Learning (ML)} models. These techniques are interesting because of the challenges in \emph{training} and \emph{evaluation} of the models in a responsible way, and because they usually produce probabilistic outputs. This means that these techniques rather than producing a single-point prediction in their output, they predict a \emph{probability distribution} over all possible outcomes. %\mscom{The next sentence probably has to stay here as we don't have the taxonomy in the intro yet}
For example, in Analysis or Synthesis this probability could be seen as the \emph{confidence} of the model in its classifications or predictions and in Generation as the probability of the next word prediction given some input text called the \emph{context}.

We map each identified AI tool to CJ system stages and figure out its classification according to the taxonomy. Each tool constitutes a single record in our repository. Each record contains the following fields:

\begin{itemize}
    \item \textbf{Name of tool.}
    \item \textbf{Purpose.} A concise statement of the tool’s intended purpose.
    \item \textbf{Description.} A fuller explanation of the tool’s operation, scope, and functionality.
    \item \textbf{Users.} The organisations deploying or trialling the tool (e.g., individual police forces, government bodies). A single tool may have multiple users.
    \item \textbf{Deployment Stage.} One of the following: Trialled, Deployed, Experimental, Stage unknown. A tool may have a different deployment stage for each identified user.
    \item \textbf{Development Type.} One of the following: In-house, Third-party, Academic collaboration. The \emph{third-party} tools constitute commercial products and are developed and provided by third party companies. The \emph{academic} development type refers to tools being the result of a collaboration between a police force and an academic institution.
    \item \textbf{Developer/Vendor.} The entity or entities responsible for developing the tool (e.g., commercial vendor, police force, academic unit).
    \item \textbf{Stage of Criminal Justice.} The stages of the criminal justice process to which the tool applies. Multiple stages may be assigned.
    \item \textbf{Classification.} Taxonomy descriptors as defined in the previous section. Tools may carry multiple descriptors where they involve multiple modalities or functionalities (e.g. Analysis (audio $\rightarrow$ text), Analysis (text $\rightarrow$ text), Synthesis ($enum^n$ $\rightarrow$ enum)).
    \item \textbf{Tool Description.} A brief description of the tool’s function corresponding to each taxonomy descriptor.
    \item \textbf{Resources.} A collection of links to publicly available online sources providing further information about the tool.
\end{itemize}

\subsection{Data collection}
This section presents how we collect the  data.   

\subsubsection{Online resources}
We rely on reliable online resources to retrieve information for our mapping. For example, good sources we have used so far are the websites of the UK Police STAR innovation fund (a UK-based fund aiming at local policing) \cite{STAR}, College of Policing, National Police Chief’s Council (NPCC), Office of the Police Chief Scientific Adviser (OPCSA) (\url{https://science.police.uk/}), Police Forces, companies that develop the tools, Universities that have collaborated with the police to develop and/or evaluate tools, the \emph{gov.uk} domain.

\subsubsection{Interviews}
We conducted semi-structured interviews with key stakeholders with the purpose to verify the data we already had, fill any gaps in our mapping, or learn about AI tools that we did not know about and had been trialled or were in use. We spoke to several police forces and policing, government and legal bodies trying to cover the whole spectrum of functions in the CJ system.  The project interviews were conducted on the basis that all data collected would be anonymised. But we have conducted 16 unique interviews and 3 follow-up interviews due to updates on AI tool since first interview. We interviewed 9 police officers across 7 police forces, 2 legal practitioners, 2 industry practitioners, and 3 regulators or oversight bodies.

%\etcom{The information above reflects the current state of the mapping where the interviews conducted in November and December 2025 have not been integrated.}

% So far, we have spoken to people from several police forces and policing, government and legal bodies in the UK:

% \begin{itemize}
%     \item National Police Chief’s Council (NPCC)
%     \item Home Office
%     \item College of Policing
%     \item British Transport Police
%     \item West Midlands Police
%     \item Hertfordshire Police
%     \item Humberside Police
%     \item Thames Valley Police
%     \item Police Scotland
%     \item Kent Police
% \end{itemize}

\subsection{An instance of the map}
\begin{figure*}[!th]
\centering
\subfloat{\includegraphics[width=0.45\textwidth]{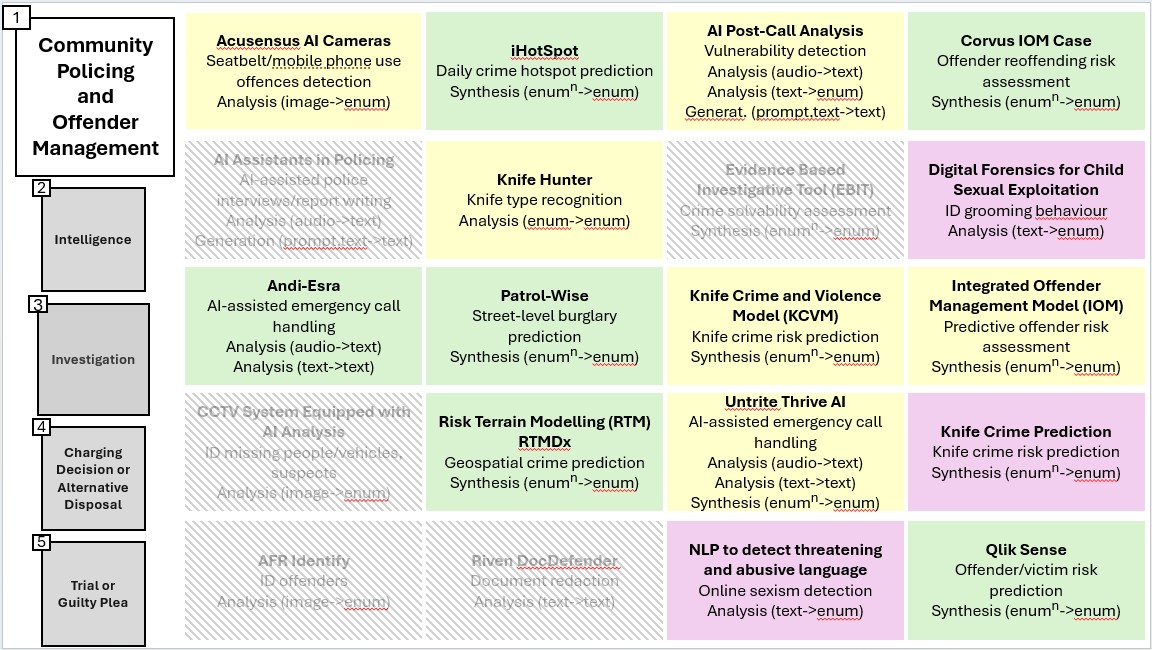}}\quad
\subfloat{\includegraphics[width=0.45\textwidth]{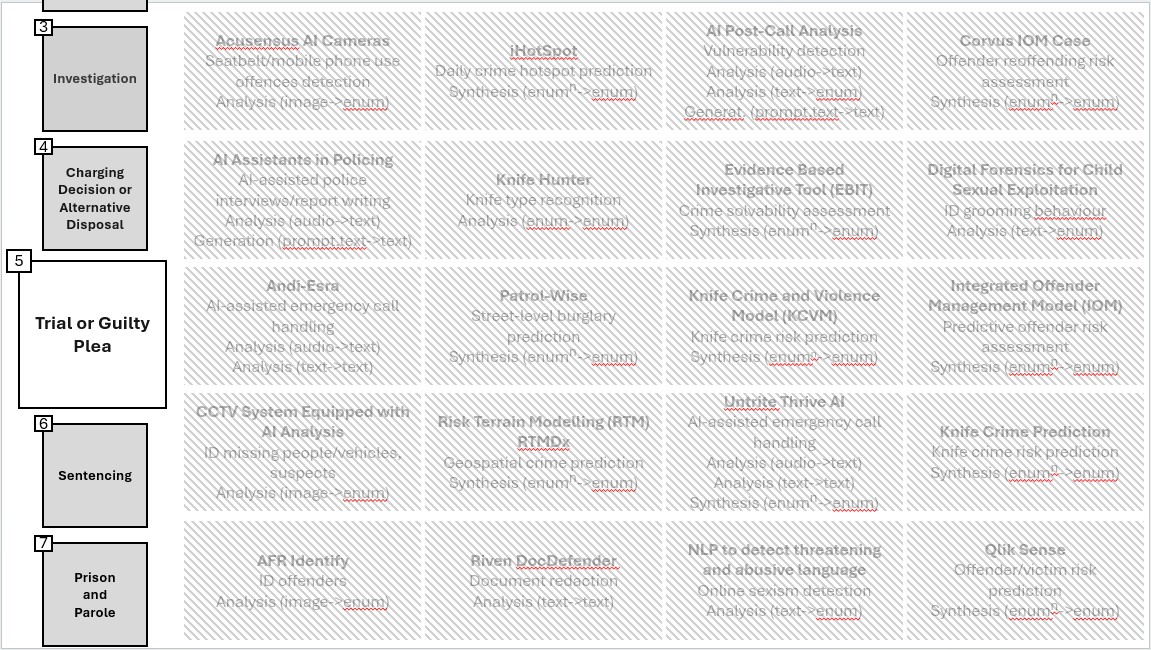}}\\

\subfloat{\includegraphics[width=0.45\textwidth]{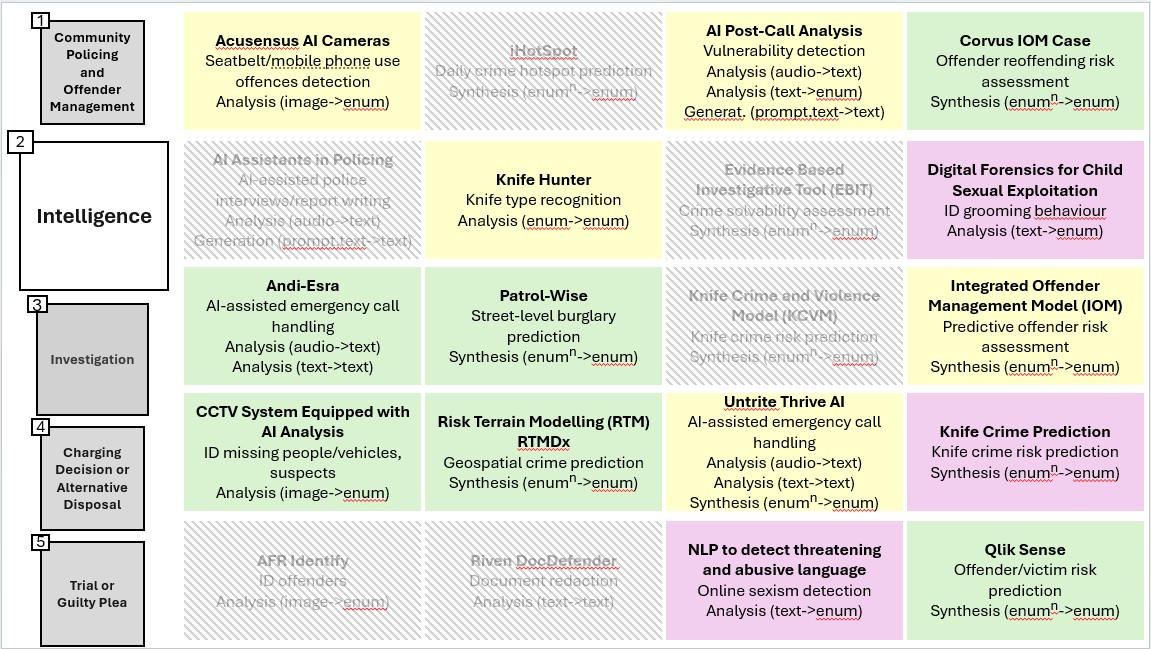}}\quad
\subfloat{\includegraphics[width=0.45\textwidth]{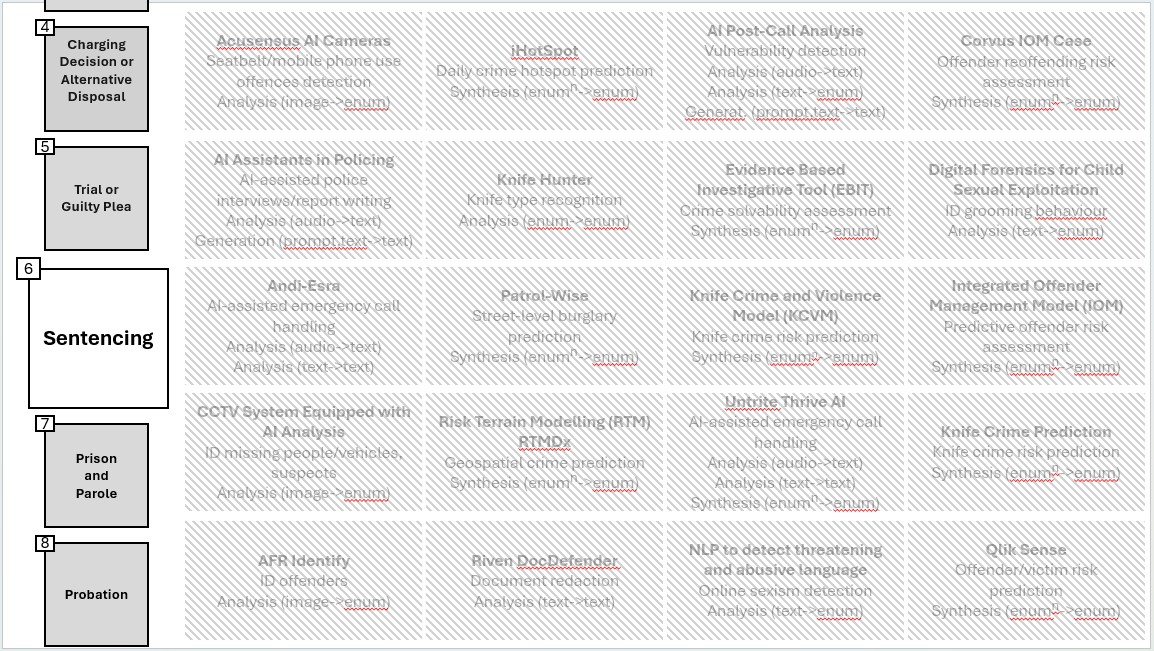}}\\

\subfloat{\includegraphics[width=0.45\textwidth]{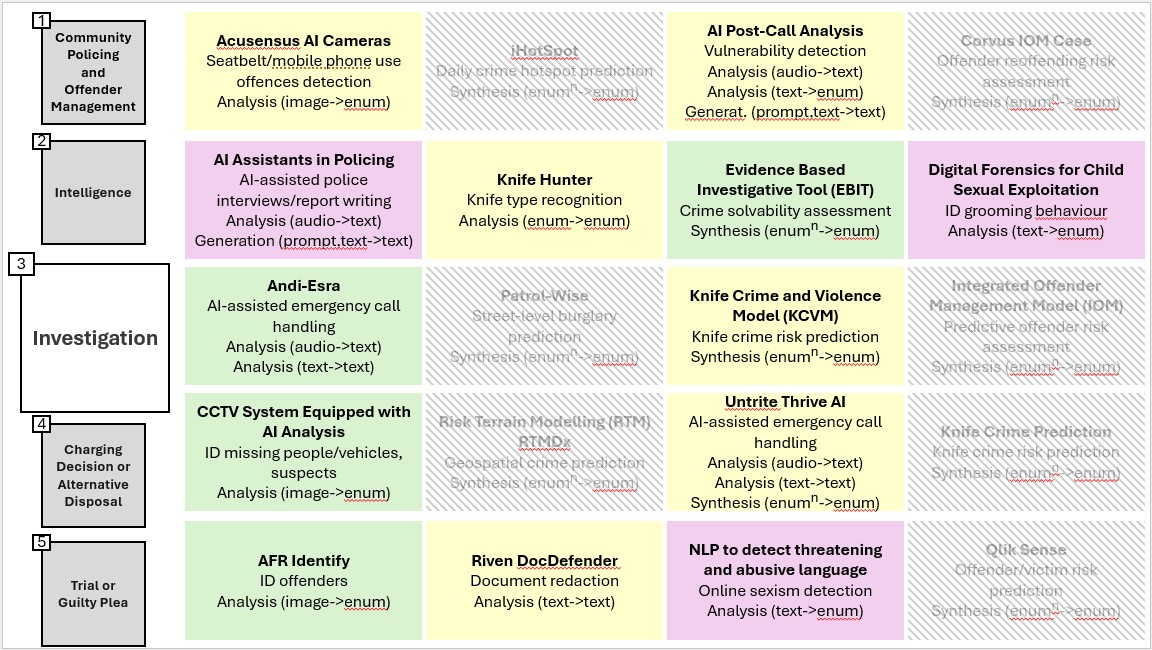}}\quad
\subfloat{\includegraphics[width=0.45\textwidth]{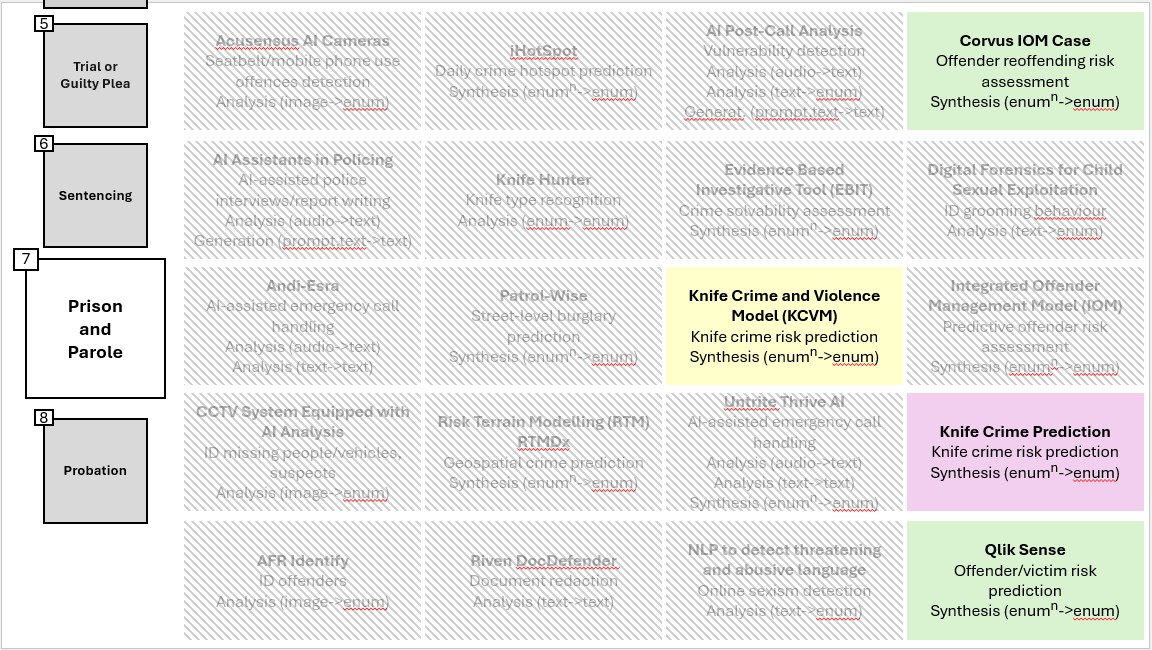}}\\

\subfloat{\includegraphics[width=0.45\textwidth]{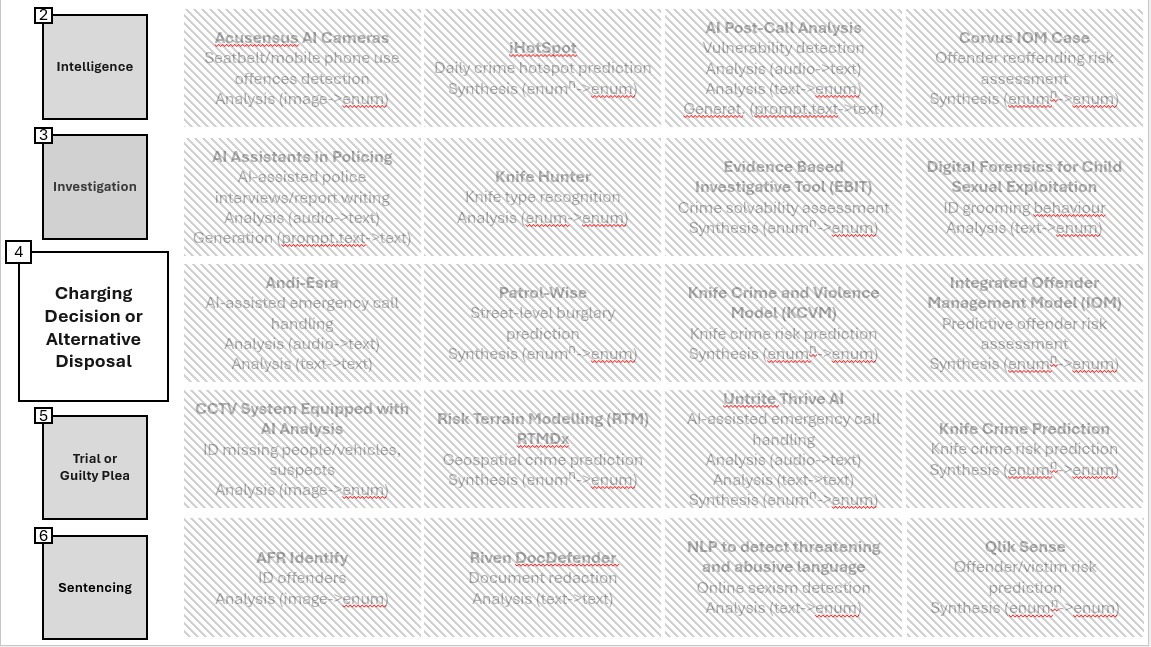}}\quad
\subfloat{\includegraphics[width=0.45\textwidth]{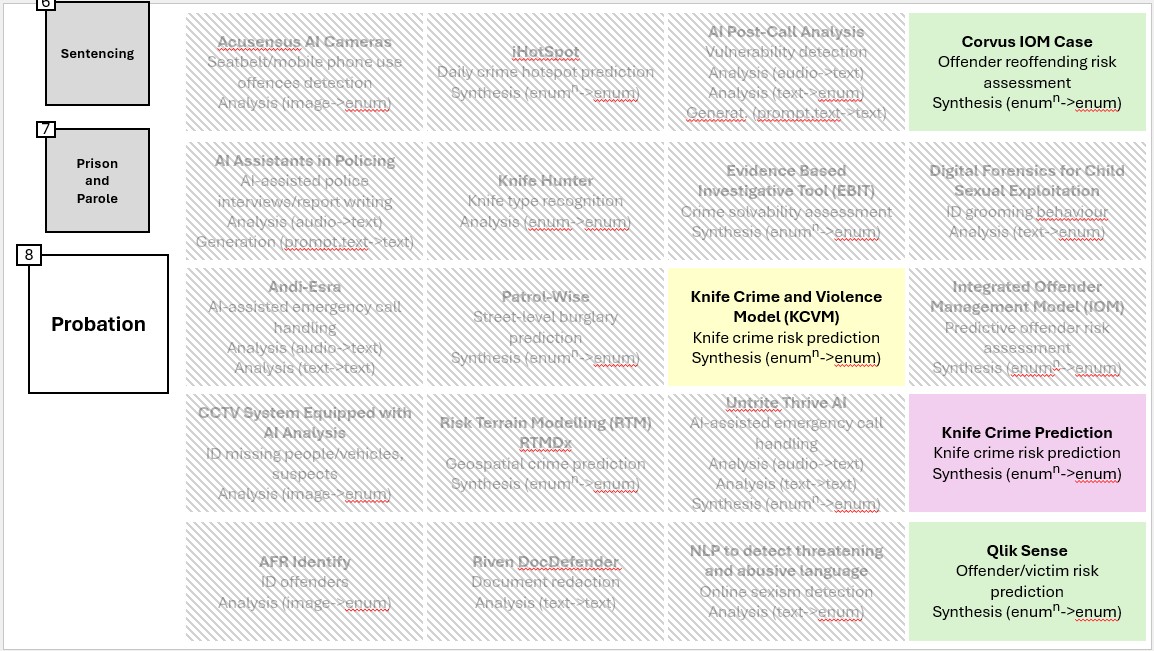}}\\
  
  \caption{Sample of AI tools currently used in the Criminal Justice (CJ) system in England and Wales. The purpose and taxonomy of each tool are stated below the name of each tool. The colour of each tool's box indicates the deployment stage of the tool: green for deployed, yellow for trialled, and pink for experimental. Shaded boxes indicate tools not used in the corresponding stages of CJ.}
  \Description{The figure presents 20 AI tools used in the Intelligence stage of the Criminal Justice process in a 5x4 grid.}
  \label{fig:mapping}
\end{figure*}

Fig.~\ref{fig:mapping} aims to illustrate our mapping. We present a sample of the AI tools in our mapping in a $5x4$ grid. Each tool is represented by a box coloured according to the deployment stage: pink for experimental, yellow for trialled, and green for deployed tools. Each box has a header at the top stating the name of the tool. A short text stating the purpose of the tool follows, and finally any identified taxonomy for this tool is shown at the bottom of the box. The gray-shaded boxes indicate tools that are not used in the corresponding CJ stages. A running ribbon on the left of the grid shows the CJ stages represented by gray boxes with the bigger white box standing out to indicate the current CJ stage. 

\subsection{Research Questions (RQs)}
In our effort to acquire a better understanding of the AI ecosystem in the CJ system in England and Wales we seek to investigate the following research questions:
\begin{itemize}
    \item[RQ1] To what extent is AI used across the CJ system in England and Wales?    
    \item[RQ2] How common is each development type (third party, in-house, academic collaboration)?
    \item[RQ3] How common is each deployment stage (experimental, trialled, deployed)?
    \item[RQ4] How common is each mode of information inference (Analysis, Synthesis, Generation)?
    \item[RQ5] What emerging technologies gain ground and need attention?
\end{itemize}

We will analyse the data in our mapping repository to respond to these questions.

\subsection{Limitations}
Our mapping is not exhaustive. The number of interviews we could conduct was limited, some of the interview candidates refused or ignored our interview invitation and a couple of them withdrew. Thus, many police forces have not been represented and other stakeholders such as industries have low representation. Apart from the interviews, our knowledge about the AI tools used across the CJ stages is limited by what is published. It can be that there are many more AI tools currently used that we do not know about or our understanding of how some tools work or in which stages in CJ is limited. This could have affected our findings in the investigated research questions. Regarding the mapping of AI tools to CJ stages, we did our best to assign tools to CJ stages by utilising any available evidence of their deployment in the corresponding CJ stages or if this was not possible due to limited or missing information, we did the assignment to the best of our understanding of where these tools could be most possibly used based on who uses them and how they work.

\section{Findings}

We identified 58 AI tools used across the CJ system in England and Wales. This list of tools is evolving and we will continue updating it with new AI tools or with status changes for existing tools. %The police forces using these tools include Avon and Somerset, Bedfordshire, Cheshire, Cumbria, Devon and Cornwall, Essex, Greater Manchester, Gwent, Hampshire, Hertfordshire, Humberside, Kent, Lincolnshire, Merseyside,  Metropolitan Police, Norfolk and Suffolk, South Wales, South Yorkshire, Surrey and Sussex, Thames Valley, West Mercia, West Midlands, and West Yorkshire Police. 

\vspace{0.1cm}
\noindent \textbf{RQ1. To what extent is AI used across the CJ system in England and Wales?}
Table~\ref{tab:analytics} presents the AI tools counts across the CJ stages. The most populated stage is \emph{Community Policing and Offender Management} (stage 1), followed by \emph{Investigation} (stage 3) and \emph{Intelligence} (stage 2). The map gets sparser when the case starts moving to the courts, namely at CJ stages 4-8. This observation may reflect greater caution with the use of AI as we move closer to the decision-making stages in the CJ system.

\begin{table}[!h]
\caption{CJ stage-specific AI tool counts from the total of 58 AI tools included in our mapping. We present the \emph{total number} of tools used at each stage, and a breakdown by \emph{inference mode} and \emph{deployment stage}.}

\resizebox{\linewidth}{!}{%
\begin{tabular}{@{}rcccccccc@{}}
\toprule
\bf \makecell[r]{} &%Number of\\ tools\textbf{}
  \bf Stage 1 &
  \bf Stage 2 &
  \bf Stage 3 &
  \bf Stage 4 &
  \bf Stage 5 &
  \bf Stage 6 &
  \bf Stage 7 &
 \bf Stage 8 \\ 

\midrule
 Total Number &\textbf{44}&37&43&2&1&0&6&6\\

 \hline
 
Analysis&25&25&33&2&1&0&1&1\\
Synthesis&18&12&7&0&0&0&6&6\\
Generation&11&9&13&1&1&0&1&1\\

\hline

Deployed&14&11&12&0&0&0&2&2\\
Trialled&17&15&19&2&1&0&1&1\\
Experimental&5&5&5&0&0&0&1&1\\
Stage Unknown&8&6&6&0&0&0&2&2\\
\bottomrule
\end{tabular}
}
\label{tab:analytics}
\end{table} 

\vspace{0.1cm}
\noindent \textbf{RQ2. How common is each development type (third party, in-house, academic collaboration)?}
We found that the third party tools is the most common development type with 57\% of the tools in our map to constitute a commercial product developed by a third party company, 22\% to have been developed in-house, and 9\% to have resulted from an academic collaboration. We observe a strong reliance on third party companies for the provision of AI tools in this domain. The development, data and algorithm hosting, maintenance, update and evaluation of software usually require many resources and specialized capabilities that the individual police forces or other involved bodies in the CJ system may not possess. Thus, they often turn to third party companies or academic institutions. Nevertheless, we still observe a considerable proportion of in-house developed tools. This development type could be preferred by police forces that acquire the necessary resources to avoid complications with the procurement process, security and data privacy concerns, and the higher costs of commercial solutions.

\vspace{0.1cm}
\noindent \textbf{RQ3. How common is each deployment stage (experimental, trialled, deployed)?}
The deployed tools represent the 33\% of all tools in our mapping repository, while there are more tools that have been trialled, i.e. 38\%. The experimental tools represent the 12\% of the tools. There is also a 16\% of tools with an unknown deployment stage. At the level of individual CJ stages, most of the tools in stages 1-3 are either deployed or have been trialled, there are only 5 experimental tools in each stage and 8-6 tools with unknown stage. All tools in stage 4 and 5 have been trialled. The tools in the other stages are well distributed across the categories of deployment stage. This data indicate that there is interest in trialling and adopting more AI tools especially in the initial stages of the CJ system. The trialled tools in these stages are more than those deployed and it is expected that many of them once they go through the proof of concept phase will be integrated in the policing processes.

\vspace{0.1cm}
\noindent \textbf{RQ4. How common is each mode of information inference (Analysis, Synthesis, Generation)?}
We found that overall 64\% of the tools in our mapping use Analysis, 33\% use Synthesis, and 26\% use Generation. The initial stages (1-3) are characterised by Analysis and/or Synthesis as the main modes of inference. Stages 7-8 are mainly characterised by Synthesis. Generation appears to be more common in the Community Policing and Offender Management (stage 1) and the Investigation stage (stage~3) with tools used for witness statements or crime reports writing, notes taking, summarization, redaction,  or retrieval of information from knowledge storages. In the earlier stages (1-3) we mainly find tools for hot-spot prediction, face recognition, emergency call handling, while in the later stages (7-8) we mainly find risk prediction tools.

\vspace{0.1cm}
\noindent \textbf{RQ5. What emerging technologies gain ground and need attention?} We found that Generation-based technologies have started being introduced or trialled in the domain of policing and law enforcement. Our interviews have revealed a strong and growing interest in these emerging technologies, particularly in the adoption and exploration of Large Language Models (LLMs). Trials of generative tools have been completed or are under the way and these usually concern tasks such as witness statement generation, translation, transcription, document summarization, question generation, generation of points to prove from witness statements, and key word generation. Evaluation of these technologies on the basis of the aforementioned tasks is limited (with the exception of the Hertfordshire Constabulary proof of concept {\em ADA} tool \cite{Walley}) at the moment. Reasons for this could be the lack of  ground-truth for many of these tasks (e.g., what is a good summary?), appropriate benchmarks, and the inability of the traditional Natural Language Processing (NLP) evaluation techniques to deal with LLM-specific issues such as hallucinations.

\section{Conclusions}
Using our methodology to map the probabilistic AI ecosystem in Criminal Justice, we started building a data repository for England and Wales. To demonstrate the value of this work, we presented some insights based on the data we have collected so far. Our next step is the development of a web interface for our repository -- to show the mapping data and various analytics based on it, and to allow more interactive engagement. This will become public and allow submissions of requests for changes, updates, or additions of new tools by any stakeholder. We are hoping that through this tool we will be able to increase the quality and richness of data in our mapping and provide a more accurate picture of the probabilistic AI ecosystem in England and Wales. We believe that this tool will serve as a useful reference point for all involved stakeholders or  for researchers in this domain. Emerging areas could be highlighted along with the need for coordination in terms of policies and regulations in these cases. Finally, this tool could promote transparency in terms of where and how AI is used in Criminal Justice for the sake of the general public good.

% \section{Discussion}
% Points of Discussion:
% \begin{itemize}
%     \item What is the current status of the use of AI in policing in the UK based on the available information?
%     \item What is the future of AI in policing?
%     \item How could we better support the stakeholders by mapping the AI ecosystem in policing?
%     \item What could be the impact of such a mapping? (law enforcement stakeholders increase their transparency to the public about their use of AI tools)
%     \item Limitations (e.g. available information - stress the significance of the collaboration between academics and stakeholders)
% \end{itemize}

% \section{Conclusions} 

\section*{Acknowledgments}
We thank Katherine Jones and Kyle Montague for the design of the mapping presentation in Fig.~\ref{fig:mapping}. This work was supported by the Engineering and Physical Sciences Research Council [grant number EP/Y009800/1], through funding from Responsible Ai UK (KP0003).   

\bibliographystyle{ACM-Reference-Format}
\bibliography{bibliography}

\end{document}